\documentclass[aps,preprint,nofootinbib]{revtex4}
\usepackage{eurosym}
\usepackage{amsfonts}
\usepackage{amsmath}
\usepackage{amssymb}
\usepackage{xcolor}
\usepackage{float}
\usepackage{graphicx}
\usepackage[utf8]{inputenc}
\begin{document}

\title{Energy of asymptotically AdS black holes in Exotic Massive Gravity and its log-extension}
\author{Robert B. Mann$^{1}$, Julio Oliva$^{2}$, S. N. Sajadi$^{3}$}
\affiliation{$^{1}$Department of Physics and Astronomy, University of Waterloo, Waterloo, Ontario, N2L 3G1,Canada.}
\affiliation{$^{2}$Departamento de F\'{\i}sica, Universidad de Concepci\'{o}n, Casilla,160-C, Concepci\'{o}n, Chile.}
\affiliation{$^{3}$Department of Physics, Shahid Beheshti University, G. C., Evin, Tehran 19839, Iran.\footnote{emails: rbmann@uwaterloo.ca, julioolivazapata@gmail.com, naseh.sajadi@gmail.com}}

\begin{abstract}
Recently a new three-dimensional theory of gravity, dubbed Exotic Massive Gravity, was proposed as a unitary theory both in the bulk as well as in the dual CFT. This is the second simplest example, the first being Minimal Massive Gravity. 
Since the divergence of the field equations vanishes on-shell, Exotic Massive Gravity has ``third-way consistency". Here we 
show how to compute  mass and angular momentum in this theory, and then implement our result in  various solutions, both for generic values of the couplings as well as at chiral points of the theory.  For the latter, the asymptotic AdS behaviour is relaxed and the metric acquires logarithmic terms, which may lead to a logarithmic CFT in the boundary. Remarkably, even in the presence of this relaxed asymptotic behaviour, the charges turn out to be finite, defining non-linear solutions of what could be called Exotic Log Gravity.
\end{abstract}
\maketitle

\section{Introduction}
Due to the absence of local degrees of freedom, General Relativity is a simpler theory in three than in four spacetime dimensions. In spite of this simplicity, the theory does contain black holes when
a negative cosmological constant is included \citep{BTZ}, whose entropy can be
correctly accounted for, microscopically, by counting states in a dual CFT
\citep{Strominger}. Including local degrees of freedom makes the theory closer to its
four dimensional counterpart;  this can be achieved by deforming the theory,
giving a mass to the graviton. In this manner by augmenting the
Einstein-Hilbert action with the parity odd, Lorentz-Chern-Simons term for the
Christoffel connection, one obtains a theory that propagates  a massive
graviton \citep{TMG1}-\citep{TMG2}. Almost a decade ago it was shown that this can also be achieved in a parity-even manner by
the addition to the Einstein-Hilbert action of a precise quadratic combination
of the Riemann curvature, such that the whole theory, when expanded around flat
spacetime, leads to the Fierz-Pauli equation for a massive spin 2 excitation
\citep{NMG}. Both of these theories possess asymptotically AdS black holes,
as well as black holes with more general asymptotics for particular interesting values of the couplings (see e.g. \citep{sols1}-\citep{solslast}).

Efforts to
understand, on microscopic terms, the entropy of some of these black holes in a
holographic context lead to the so-called bulk-boundary clash of these
theories, according to which positivity of the energy of the bulk massive
graviton seems to be achieved only at the cost of introducing a negative central
charge in the dual theory. A theory that avoids this problem is
Minimal Massive Gravity (MMG).  MMG supplements the field equations of
Topologically Massive Gravity (TMG) with a symmetric, rank-two tensor containing up
to second derivatives of the metric \citep{MMG}. The Lovelock theorem \citep{Lovelock} implies that
the latter tensor cannot have an identically vanishing divergence.
Nevertheless the theory is consistent in the so-called ``third-way", since the
divergence of the field equations is identically proportional to those same field
equations. Even though the field equations for MMG
 cannot be obtained from a diffeomorphism invariant action principle, a
Lagrangian formulation of the theory can be obtained by considering extra
fields, which is useful when computing the central charge of the boundary
theory.

Recently, a new theory fulfilling ``third-way" consistency was proposed  \citep{EMG}. This theory, dubbed Exotic Massive Gravity (EMG), supplements the
Einstein equations with a term that contains up to third derivatives of the
metric and is constructed with combinations and derivatives of the Cotton
tensor, which can also be added by itself with its own coupling, still
maintaining the consistency of the field equations. This theory turns out to
be the next-to-simplest theory fulfilling   third-way consistency, from an
infinite family that  is constructed from a seed  rank-two  and divergence free
symmetric tensor $\mathcal{G}_{\mu\nu}$. When $\mathcal{G}_{\mu\nu}$ is
proportional to the Einstein tensor, the theory obtained is MMG, whereas EMG is
obtained from choosing $\mathcal{G}_{\mu\nu}$ proportional to the Cotton tensor.

An exploration of the vacua of EMG was subsequently undertaken
\citep{vacuaEMG}, which includes asymptotically AdS geometries as well as
asymptotically Lifshitz black holes at a particular curve of the space of
parameters of the theory, the latter resembling the chiral point of TMG \citep{ChiralGravity}. Given these theories it would be interesting to explore their
thermodynamics. While conjugate variables such as the temperature and rotation
velocity of the black hole  can be directly computed from the
geometry, the lack of a diffeomorphism invariant action, in terms of the
metric uniquely, makes it difficult to compute the associated global charges as well as the
entropy. 

The purpose of this paper is to address this question. Using the Abbot-Deser-Tekin (ADT) approach
\citep{AD}-\citep{ADTs}, which naturally adapts to the present setup, we shall construct formulae
for the global charges of the theory. Focusing on asymptotically AdS solutions of MMG that
exist for generic values of the coupling constants, we shall apply our results to a computation of  the mass
of the rotating Banados-Teitelboim-Zanelli  (BTZ) solution, showing that the parity even terms contribute to the energy as expected, from the mass parameter in GR with an effective
Newton's constant. Furthermore, as is the case in TMG, when EMG is supplemented by  a Lorentz-Chern-Simons term, the mass of the rotating BTZ black hole receives an extra
contribution from this parity odd term, which depends on the rotation
parameter $a$. We then consider other axially symmetric solutions. At the so-called chiral point, log-deformations of the extremal BTZ black hole exist, and we show that they remarkably lead to  finite charges. We conclude with comments and possible extensions.

\section{Constructing the ADT charges in EMG}\label{sec2}

 Exotic Massive Gravity is defined by the following field equations \citep{EMG}:
\begin{equation}\label{eq1}
R_{\mu \nu}-\dfrac{1}{2}g_{\mu \nu}R+\Lambda g_{\mu \nu}+\dfrac{1}{\mu}C_{\mu \nu}-\dfrac{1}{m^{2}}H_{\mu \nu}+\dfrac{1}{m^{4}}L_{\mu \nu}= 0 \, ,
\end{equation}
where 
\begin{equation}\label{eq2}
C_{\mu \nu}=\dfrac{1}{2}\epsilon_{\mu}^{\ \alpha \beta}\nabla_{\alpha}(R_{\beta \nu}-\dfrac{1}{4}g_{\nu \beta}R)\, ,\hspace{0.5cm}H_{\mu \nu}=\epsilon_{\mu}^{\ \alpha \beta}\nabla_{\alpha}C_{\nu \beta}\, ,\hspace{0.5cm}L_{\mu \nu}=\dfrac{1}{2}\epsilon_{\mu}^{\ \alpha \beta}\epsilon_{\nu}^{\ \gamma \sigma}C_{\alpha \gamma}C_{\beta \sigma}\, ,
\end{equation}
and $C_{\mu \nu} $ is the Cotton tensor. 

It is useful to define $ \Lambda=-\dfrac{1}{l^{2}} $,  $l$ being the curvature radius of the maximally symmetric AdS solution of the field equations  (\ref{eq1}).

To follow the ADT approach, we linearize the field equations (\ref{eq1}) around a locally AdS$_{3}$ vacuum (not necessarily global AdS) of radius $l$, by decomposing the metric as:
\begin{equation}
g_{\mu \nu}=\bar{g}_{\mu \nu}+h_{\mu \nu}\, ,
\end{equation}  
where the background metric $ \bar{g}_{\mu \nu} $ satisfies  
\begin{equation}
\bar{R}_{\mu \nu}+\dfrac{2}{l^{2}}\bar{g}_{\mu \nu}=0\, ,
\end{equation}
and $ h_{\mu \nu} $ is a small perturbation. 
At the linearized level, the equation of motion (\ref{eq1}) reduces to
\begin{equation}\label{eq5}
{\cal G}^{(l)}_{\mu \nu}
+\dfrac{1}{\mu}C^{(l)}_{\mu \nu}-\dfrac{1}{m^{2}}H^{(l)}_{\mu \nu}+\dfrac{1}{m^{4}}L^{(l)}_{\mu \nu} \equiv T_{\mu \nu} \, ,
\end{equation}
where $T_{\mu \nu}$ contains the matter contributions as well as the contributions of order $\mathcal{O}(h^2)$, and
\begin{equation}
{\cal G}^{(l)}_{\mu \nu} = 
R^{(l)}_{\mu \nu}-\dfrac{1}{2}\bar{g}_{\mu \nu}R^{(l)}+\dfrac{2}{l^{2}} h_{\mu \nu}\, ,
\end{equation}
with
\begin{align}
R^{(l)}_{\mu \nu} &=\dfrac{1}{2}\left[-\bar{\nabla}^{2}h_{\mu \nu}-\bar{\nabla}_{\mu}\bar{\nabla}_{\nu}h +\bar{\nabla}_{\mu}\bar{\nabla}_{\sigma}h^{\sigma}_{\nu}+\bar{\nabla}_{\nu}\bar{\nabla}_{\sigma}h^{\sigma}_{\mu}\right]  \\
R^{(l)} &=-\bar{\nabla}^{2}h+\bar{\nabla}_{\rho}\bar{\nabla}_{\sigma}h^{\rho \sigma}+\dfrac{2}{l^{2}}h 
\end{align}
where $ h=\bar{g}^{\mu \nu} h_{\mu \nu}$. We also have
\begin{equation}
C_{\mu \nu}^{(l)}=\dfrac{1}{2}\epsilon_{\mu}^{\ \alpha \beta}\bar{\nabla}_{\alpha}(R_{\beta \nu}^{(l)}-\dfrac{1}{4}\bar{g}_{\nu \beta}R^{(l)}+\dfrac{2}{l^{2}}h_{\mu \nu}) 
\end{equation}
which is the linearization of the Cotton tensor.

Choosing the gauge $ \bar{\nabla}^{\mu}h_{\mu \nu}=\bar{\nabla}_{\nu}h $ and   making use of linearized Bianchi identity ($ \bar{\nabla}_{\mu}\bar{\nabla}_{\nu}h=-\bar{\nabla}^{2}h_{\mu \nu} $) the left-hand side of (\ref{eq5}) becomes
\begin{equation}\label{eq8}
-\bar{\nabla}^{2}h_{\mu \nu}+\dfrac{2}{l^{2}}h_{\mu \nu}+\dfrac{1}{\mu}\epsilon_{\mu}^{\ \alpha \beta}\bar{\nabla}_{\alpha}\left(-\bar{\nabla}^{2}h_{\beta \nu}+\dfrac{2}{l^{2}}h_{\beta \nu}\right)-\dfrac{1}{m^{2}}\left(-\bar{\nabla}^{4}h_{\mu \nu}+\dfrac{2}{l^{2}}\bar{\nabla}^{2}h_{\mu \nu} \right)   
\end{equation}
where the contribution from $L^{(l)}_{\mu \nu}$ vanishes since  $L_{\mu \nu}$ 
is quadratic in the Cotton tensor and we  are linearizing around a constant curvature (therefore conformally flat) background. 
Rearranging \eqref{eq8} implies that  \eqref{eq5} becomes
\begin{equation}\label{eq9}
\left(\bar{\nabla}^{2}-\dfrac{2}{l^{2}}\right)\left(h_{\mu \nu}+\dfrac{1}{\mu}\epsilon_{\mu}^{\ \alpha \beta}\bar{\nabla}_{\alpha}h_{\beta \nu}\right)-\dfrac{1}{m^{2}}\left(\bar{\nabla}^{2}-\dfrac{2}{l^{2}}\right)\bar{\nabla}^{2}h_{\mu \nu}=T_{\mu \nu}    
\end{equation}  
where $ C_{\mu \nu} $ and  $ H_{\mu \nu} $ are traceless, yielding $ h=0 $ upon taking the trace of \eqref{eq1}, in vacuum and up to order $\mathcal{O}(h^2)$. Since the linearized field equations (\ref{eq9}) are on-shell divergenceless,  their contraction with a Killing vector of the background leads to a conserved current.
 
To obtain the charge associated with a solution having  a Killing vector $ \bar{\xi}_{\nu} $, we 
contract both sides of \eqref{eq9} with $ \bar{\xi}_{\nu} $ and integrate, obtaining
\begin{equation}\label{eq10}
Q^{\mu}(\bar{\xi})= Q_E^{\mu}(\bar{\xi}) +\dfrac{1}{\mu} Q_C^{\mu}(\bar{\xi}) -\dfrac{1}{m^{2}} Q_H^{\mu}(\bar{\xi}) = 
\int d^{n-1}x \sqrt{-\bar{g}}\bar{\xi}_{\nu}T^{\mu \nu}=\int_{\Sigma}dS_{i}F^{\mu i} \, ,
\end{equation}
where $ F^{\mu \nu} $ is an anti-symmetric tensor that satisfies $ T^{\mu \nu}\bar{\xi}_{\nu}=\bar{\nabla}_{\nu}F^{\mu \nu} $. Provided one uses the on-shell consistency of the field equations, such a 2-form always exists.

The $Q_E^{\mu}$, $Q_C^{\mu}$, and $Q_H^{\mu}$ terms respectively correspond to the terms proportional
to unity, $1/\mu$ and $1/m^2$ obtained from \eqref{eq9}. Explicitly the first two of these are \citep{DTTMG} 
\begin{align}
Q_E^{\mu}(\bar{\xi})&=\dfrac{1}{8\pi G_{3}}\int dS_{i}\sqrt{-\bar{g}}( \bar{\xi}_{\nu}\bar{\nabla}^{\mu}h^{i \nu}-\bar{\xi}_{\nu}\bar{\nabla}^{i}h^{\mu \nu}+\bar{\xi}^{\mu}\bar{\nabla}^{i}h-\bar{\xi}^{i}\bar{\nabla}^{\mu}h+h^{\mu \nu}\bar{\nabla}^{i}\bar{\xi}_{\nu}-h^{i\nu}\bar{\nabla}^{\mu}\bar{\xi}_{\nu} \nonumber\\
&\qquad\qquad\qquad\qquad\qquad +\bar{\xi}^{i}\bar{\nabla}_{\nu}h^{\mu \nu}-\bar{\xi}^{\mu}\bar{\nabla}_{\nu}h^{i\nu}+h\bar{\nabla}^{\mu}\bar{\xi}^{i}) 
\label{QE} \\
Q_C^{\mu}(\bar{\xi}) &=\dfrac{1}{8\pi G_{3}}\int dS_{i}\sqrt{-\bar{g}}(\epsilon^{\mu i \beta}{\cal G}^{(l)}_{\nu \beta}\bar{\xi}^{\nu}+\epsilon^{\nu i \beta}{\cal G}^{(l)\mu}{}_{ \beta}\bar{\xi}_{\nu}+\epsilon^{\mu \nu \beta}{\cal G}^{i}{}_{\beta}\bar{\xi}_{\nu})+Q^{\mu}_{E}(\epsilon \bar{\nabla}\bar{\xi}). 
\label{QC} 
\end{align}
The remaining task is to compute $Q_H^{\mu}(\bar{\xi})$, the contribution to the conserved charges coming from the $H$ term.  We begin
by noting that 
\begin{equation}
H^{\mu \nu}=\epsilon^{\mu \rho \sigma}\nabla_{\rho}C_{\sigma}^{\nu}=\dfrac{1}{2}\left[ \epsilon^{\mu \rho \sigma}\nabla_{\rho}C_{\sigma}^{\nu}+\epsilon^{\nu \rho \sigma}\nabla_{\rho}C^{\mu}_{\sigma}\right] \, ,
\end{equation} 
whose linearization yields
\begin{equation}\label{eq15}
\sqrt{-\bar{g}}H^{(l)\mu \nu}\bar{\xi}_{\nu}=\dfrac{\sqrt{-\bar{g}}}{2}\left[\epsilon^{\mu \rho \sigma}(\bar{\nabla}_{\rho}C^{(l)\nu}_{\sigma})\bar{\xi}_{\nu}+\epsilon^{\nu \rho \sigma}(\bar{\nabla}_{\rho}C^{(l)\mu}_{\sigma})\bar{\xi}_{\nu} \right]\, . 
\end{equation}

We now need to move the Killing vector inside the covariant derivatives, using
\begin{align}\label{eq17}
\sqrt{-\bar{g}}\bar{\nabla}_{\rho}\left( \epsilon^{\mu \rho \sigma}C^{(l)\nu}_{\sigma}\right)\bar{\xi}_{\nu} &=\bar{\nabla}_{\rho}\left(\sqrt{-\bar{g}}\epsilon^{\mu \rho \sigma}C^{(l)\nu}_{\sigma}\bar{\xi}_{\nu}\right)-\sqrt{-\bar{g}}\epsilon^{\mu \rho \sigma} C^{(l)\nu}_{\sigma}\bar{\nabla}_{\rho}\bar{\xi}_{\nu}  \\
\sqrt{-\bar{g}}\bar{\nabla}_{\rho}\left( \epsilon^{\nu \rho \sigma}C^{(l)\mu}_{\sigma}\right)\bar{\xi}_{\nu} &=\bar{\nabla}_{\rho}\left(\sqrt{-\bar{g}}\epsilon^{\nu \rho \sigma}C^{(l)\mu}_{\sigma}\bar{\xi}_{\nu}\right)-\sqrt{-\bar{g}}\epsilon^{\nu \rho \sigma} C^{(l)\mu}_{\sigma}\bar{\nabla}_{\rho}\bar{\xi}_{\nu}
\end{align}
yielding
\begin{align}
&\bar{\nabla}_{\rho}\left[\sqrt{-\bar{g}}\epsilon^{\mu \rho}{}_{\sigma}C^{(l)\nu \sigma}\bar{\xi}_{\nu}+\sqrt{-\bar{g}}\epsilon^{\nu \rho}{}_{\sigma}C^{(l)\mu \sigma}\bar{\xi}_{\nu}+\sqrt{-\bar{g}}\epsilon^{\mu \nu}{}_{\sigma}C^{(l)\rho \sigma}\bar{\xi}_{\nu}\right] \nonumber \\
& =  \sqrt{-\bar{g}}\left[\epsilon^{\mu \rho \sigma}(\bar{\nabla}_{\rho}C^{(l)\nu}_{\sigma}) +\epsilon^{\nu \rho \sigma}(\bar{\nabla}_{\rho}C^{(l)\mu}_{\sigma})  \right]\bar{\xi}_{\nu} + \sqrt{-\bar{g}}\left[\epsilon^{\mu \rho}{}_{\sigma}C^{(l)\nu \sigma}
+\epsilon^{\mu \nu}{}_{\sigma}C^{(l)\rho \sigma} \right]\bar{\nabla}_{\rho}\bar{\xi}_\nu \nonumber \\
&\qquad + \sqrt{-\bar{g}}\epsilon^{\mu \nu}{}_{\sigma} \bar{\nabla}_{\rho}C^{(l)\rho \sigma}\bar{\xi}_{\nu}
+ \sqrt{-\bar{g}}\epsilon^{\nu \rho}{}_{\sigma}C^{(l)\mu \sigma}\bar{\nabla}_{\rho}\bar{\xi}_{\nu} \, , \nonumber \\
& = 2 \sqrt{-\bar{g}}H^{(l)\mu \nu}\bar{\xi}_{\nu} + \sqrt{-\bar{g}} \bar{X}_\sigma C^{(l)\mu \sigma}\, , 
\end{align}
where $\bar{\nabla}_{\rho}C^{(l)\rho \sigma} = 0$ and the second term in the middle line vanishes because 
of the antisymmetry of  $\bar{\nabla}_{\rho}\bar{\xi}_\nu$.  Noting that 
$\bar{X}_\sigma = \epsilon^{\nu \rho}{}_{\sigma} \bar{\nabla}_{\rho}\bar{\xi}_{\nu}$ is also a Killing field
(the dualized Killing vector) in the locally AdS background, we obtain
\begin{equation}\label{eq19}
2 \sqrt{-\bar{g}}H^{(l)\mu \nu}\bar{\xi}_{\nu}=\bar{\nabla}_{\rho}\left[\sqrt{-\bar{g}}\epsilon^{\mu \rho}{}_{\sigma}C^{(l)\nu \sigma}\bar{\xi}_{\nu}+\sqrt{-\bar{g}}\epsilon^{\nu \rho}{}_{\sigma}C^{(l)\mu \sigma}\bar{\xi}_{\nu}+\sqrt{-\bar{g}}\epsilon^{\mu \nu}{}_{\sigma}C^{(l)\rho \sigma}\bar{\xi}_{\nu}\right]-\sqrt{-\bar{g}} \bar{X}^{\sigma}C^{(l)\mu}_{\sigma}
\end{equation} 
where the last term can be written as a surface integral using the same logic as in TMG. 

Hence the  contribution to the energy coming from the tensor $H^{(l)\mu \nu}$ is
\begin{align}
&Q_H^{\mu}(\bar{\xi})= \dfrac{1}{8 \pi G_{3}}\int 2\sqrt{-\bar{g}}H^{(l)\mu \nu }\bar{\xi}_{\nu}d^{3}x \nonumber\\
&=\dfrac{1}{8\pi G_{3}}\left[\int_{\Sigma}\left[\sqrt{-\bar{g}}\epsilon^{\mu i}{}_{\sigma}C^{(l)\nu \sigma}+\sqrt{-\bar{g}}\epsilon^{\nu i}{}_{\sigma}C^{(l)\mu \sigma} +\sqrt{-\bar{g}}\epsilon^{\mu \nu}{}_{\sigma}C^{(l)i \sigma} \right] \bar{\xi}_{\nu} dS_{i} - \int\left(\sqrt{-\bar{g}} \bar{X}^{\sigma}C^{(l)\mu}_{\sigma}\right)d^{3}x\right]  
\label{eq20}
\end{align}

 Let's now look at the last term of equation (\ref{eq20}).
Writing $ C_{\mu \nu} $ in its symmetric form we have
\begin{equation}\label{eq21}
2\sqrt{-\bar{g}}\bar{X_{\nu}}C^{(l)\mu \nu}=\bar{X}_{\nu}\left( \epsilon^{\mu \alpha}{}_{\beta}\bar{\nabla}_{\alpha}{\cal G}^{(l)\nu \beta}+\epsilon^{\nu \alpha}{}_{\beta}\bar{\nabla}_{\alpha}{\cal G}^{(l)\mu \beta}\right) 
\end{equation} 
and using 
\begin{equation}\label{eq22}
\epsilon^{\mu \alpha}{}_{\beta}\bar{X}_{\nu} \bar{\nabla}_{\alpha}{\cal G}^{(l)\nu \beta}=\bar{\nabla}_{\alpha}\left[ \epsilon^{\mu \alpha}{}_{\beta}\bar{X}_{\nu}{\cal G}^{(l)\nu \beta} \right]-\epsilon^{\mu \alpha}{}_{\beta}\bar{\nabla}_{\alpha}\bar{X}_{\nu}{\cal G}^{(l)\nu \beta}
\end{equation}
we obtain  
\begin{multline}\label{eq25}
\bar{\nabla}_{\alpha}\left[ \epsilon^{\mu \alpha}{}_{\beta}\bar{X}_{\nu}{\cal G}^{(l)\nu \beta}+ \epsilon^{\nu \alpha}{}_{\beta}\bar{X}_{\nu}{\cal G}^{(l)\mu \beta}+\epsilon^{\mu \nu \beta}\bar{X}_{\nu}{\cal G}^{(l)\alpha}_{\beta}\right] 
\\
=
\epsilon^{\mu \alpha}{}_{\beta}\bar{X}_{\nu} \bar{\nabla}_{\alpha}{\cal G}^{(l)\nu \beta}+\epsilon^{\nu \alpha}{}_{\beta}\bar{X}_{\nu} \bar{\nabla}_{\alpha}{\cal G}^{(l)\mu \beta}+\epsilon^{\mu \nu \beta}\bar{X}_{\nu} \bar{\nabla}_{\alpha}{\cal G}^{(l)\alpha}_{\beta}
\\+\left(\epsilon^{\mu \alpha}{}_{\beta}\bar{\nabla}_{\alpha}\bar{X}_{\nu}{\cal G}^{(l)\nu \beta}+ \epsilon^{\nu \alpha}{}_{\beta}\bar{\nabla}_{\alpha}\bar{X}_{\nu}{\cal G}^{(l)\mu \beta}+\epsilon^{\mu \nu \beta}\bar{\nabla}_{\alpha}\bar{X}_{\nu}{\cal G}^{(l)\alpha}_{\beta}\right) \\
= \epsilon^{\mu \alpha}{}_{\beta}\bar{X}_{\nu} \bar{\nabla}_{\alpha}{\cal G}^{(l)\nu \beta}+\epsilon^{\nu \alpha}{}_{\beta}\bar{X}_{\nu} \bar{\nabla}_{\alpha}{\cal G}^{(l)\mu \beta}+\epsilon^{\mu \nu \beta}\bar{X}_{\nu} \bar{\nabla}_{\alpha}{\cal G}^{(l)\alpha}_{\beta}  + \epsilon^{\nu \alpha}{}_{\beta}\bar{\nabla}_{\alpha}\bar{X}_{\nu}{\cal G}^{(l)\mu \beta}\, ,
\end{multline} 
where the first and last terms of the middle line in brackets cancel because $\bar{X}_\nu$ is a Killing vector.
Recalling that $\bar{X}_\sigma = \epsilon^{\nu \rho}{}_\sigma \bar{\nabla}_{\rho}\bar{\xi}_{\nu}$   and $\bar{\nabla}_{\rho}\bar{\nabla}_{\mu}\bar{\xi}_{\nu}=\bar{R}_{\nu \mu \rho}{}^{\sigma}\bar{\xi}_{\sigma}$, we get
\begin{equation}
\epsilon^{\nu \alpha}{}_{\beta}\bar{\nabla}_{\alpha}\bar{X}_{\nu}{\cal G}^{(l)\mu \beta}
=\epsilon^{\nu \alpha}{}_{\beta}\epsilon^{\rho \sigma}{}_{\nu}\bar{\nabla}_{\alpha}\bar{\nabla}_{\rho}\bar{\xi}_{\sigma}{\cal G}^{(l)\mu \beta} 
=\epsilon^{\nu \alpha}{}_{\beta}\epsilon^{\rho \sigma}{}_{\nu}\left(\bar{R}_{\sigma \rho \alpha}{}^{\gamma}\bar{\xi}_{\gamma}\right){\cal G}^{(l)\mu \beta} =-4\Lambda \bar{\xi}_{\beta} {\cal G}^{(l)\mu \beta}
\end{equation}
where the last relation follows from
\begin{equation}
\bar{R}_{\mu \sigma \nu \rho}=\Lambda\left( \bar{g}_{\mu \nu}\bar{g}_{\sigma \rho}-\bar{g}_{\mu \rho}\bar{g}_{\sigma \nu}\right)
\end{equation}
which is the assumption of a constant curvature background.
Consequently the last term in (\ref{eq20}) contributes as
\begin{align}\label{eq30}
2\int \sqrt{-\bar{g}}\bar{X}_{\nu}C^{(l)\mu \nu}d^{3}x&=\int \bar{\nabla}_{\alpha}\left[\epsilon^{\nu \alpha}{}_{\beta} \bar{X}_{\nu}{\cal G}^{(l)\mu \beta}+\epsilon^{\mu \alpha}{}_{\beta}\bar{X}_{\nu}{\cal G}^{(l)\nu \beta}+\epsilon^{\mu \nu \beta}\bar{X}_{\nu}{\cal G}^{(l)\alpha}_{\beta}\right]d^{3}x \nonumber\\ 
&\qquad - 4 \Lambda \int \bar{\xi}_{\beta}{\cal G}^{(l)\mu \beta}d^{3}x
\end{align}

The final result of the new contribution to the Abbott-Deser-Tekin charges is
\begin{align}
Q_H^{\mu}(\bar{\xi}) &= \dfrac{1}{8 \pi G_{3}}\int 2\sqrt{-\bar{g}}H^{(l)\mu \nu }\bar{\xi}_{\nu}d^{3}x \nonumber\\
&=\dfrac{1}{8\pi G_{3}}\int_{\Sigma}\left[\sqrt{-\bar{g}}\epsilon^{\mu i}{}_{\sigma}C^{(l)\nu \sigma}+\sqrt{-\bar{g}}\epsilon^{\nu i}{}_{\sigma}C^{(l)\mu \sigma} +\sqrt{-\bar{g}}\epsilon^{\mu \nu}{}_{\sigma}C^{(l)i \sigma} \right] \bar{\xi}_{\nu} dS_{i} \nonumber\\
&\quad - \dfrac{1}{16 \pi G_{3}}\int \left[\epsilon^{\nu i}{}_{\beta}{\cal G}^{(l)\mu \beta}+\epsilon^{\mu i}{}_{\beta}{\cal G}^{(l)\nu \beta}+\epsilon^{\mu \nu \beta}{\cal G}^{(l)i}_{\beta}\right]\bar{X}_{\nu}dS_{i} +\dfrac{ \Lambda}{4 \pi G_{3}}\int \bar{\xi}_{\beta}{\cal G}^{(l)\mu \beta}d^{3}x 
\label{eq31}
\end{align}
where the last term of equation (\ref{eq31}) is the Einstein surface term ($ Q^{\mu}_{E}(\bar{\xi}) $). 

Using   (\ref{QE}), (\ref{QC}) and (\ref{eq31}) the desired gauge invariant conserved charges for a maximally symmetric background in EMG is:
\begin{equation}\label{chargesfinal}
Q^{\mu}(\bar{\xi})=Q^{\mu}_{E}(\bar{\xi})+\dfrac{1}{\mu}Q_{C}^{\mu}(\bar{\xi})-\dfrac{1}{m^{2}}Q^{\mu}_{H}(\bar{\xi}). 
\end{equation} 
This expression will allow us to compute the charges in the following sections, for solutions that are deformations of constant curvature background, a family that includes rotating black holes, as well as log-deformations of extremal BTZ.

\section{Mass and angular momentum for axisymmetric vacua of EMG: Generic couplings}

We now apply our result \eqref{chargesfinal} to a variety of interesting solutions to EMG.

\textbf{BTZ spacetime}\\
The BTZ metric 
\begin{equation}\label{BTZmetric}
ds^{2}=-N^{2}(r)dt^{2}+\dfrac{dr^{2}}{N^{2}(r)}+r^{2}\left(N^{\phi}dt+d\phi\right)^{2} 
\end{equation}
with
\begin{equation}
N^{2}(r)=-M+\dfrac{r^{2}}{l^{2}}+\dfrac{a^{2}}{4r^{2}}\hspace{1cm},\hspace{1cm}N^{\phi}(r)=-\dfrac{a}{2r^{2}}
\end{equation}
is also a solution of the EMG field equations.   Even though the tensors $C_{\mu\nu}$ and $H_{\mu\nu}$ vanish identically,  we will see that they do contribute non-trivially to the charges.  

As in GR, to define the constant curvature background, let us set $M=a=0$ so that 
\begin{equation}\label{eq11}
ds^{2}_{BG}=-\dfrac{r^{2}}{l^{2}}dt^{2}+\dfrac{l^{2}}{r^2}dr^{2}+r^{2}d\phi^{2} \, .
\end{equation} 
is  the background metric.  The
BTZ metric can then be regarded as a perturbation from this background by setting  
\begin{equation}\label{eq12}
ds_{p}^{2}=M dt^{2}+\dfrac{M l^{4}}{r^{4}}dr^{2}-adt d\phi\, ,
\end{equation}
where we have kept the leading order in the radial expansion as $r\rightarrow+\infty$. The energy, obtained as the conserved charge associated with future-directed time translations generated by the Killing vector $\bar{\xi}^{\mu}=(-1,0,0)$, finally reads

\begin{equation}\label{eq13}
E= Q^{\mu}(\bar{\xi}) =M-{\dfrac{a}{{l}^{2}\mu}}-\dfrac{M}{m^2 l^2} 
\end{equation}
using \eqref{chargesfinal}.

Note that in the limit of large $m$ and large $\mu$ we obtain the expected relation $E=M$.
The angular momentum
\begin{equation}
J= Q^{\mu}(\bar{\xi}) = a-{\dfrac {M}{\mu}}-{\dfrac {a}{{m}^{2}{l^2}}}
\end{equation}
 is likewise obtained using the Killing field $\bar{\xi}^{\mu}=\delta^{\mu}{}_{\phi} $.

We have introduced a suitable global normalization to match the charges of BTZ black hole in TMG reported in \cite{testcharges}.
It is interesting to notice that the parity-even terms in the field equations, i.e. the Einstein and $H$ tensors contribute to the energy and angular momentum by means of the mass term $M$ and rotation parameter $a$, respectively. On the other hand, the contribution of the Lorentz-Chern-Simons terms, contributes to the charges in a ``twisted fashion", as  originally observed in the presence of topological matter \cite{topomattertwisted}.
\\

 \textbf{``Static" AdS waves with fast fall-off}
 \\
 For generic values of the couplings, EMG has also been shown \cite{vacuaEMG} to
admit AdS-wave solutions, which can be written as
\begin{equation}\label{adswave}
ds^{2}=\left(-\dfrac{2\rho}{l^{2}}+F(\rho)\right)dt^{2}+\dfrac{l^{2}d\rho^{2}}{4\rho^{2}}-2lF(\rho)dt d\phi+(2\rho+l^{2}F(\rho))d\phi^{2}\, ,
\end{equation}
here 
\begin{equation}\label{eq40}
F(\rho)=a_{+}\rho^{p_{+}}+a_{-}\rho^{p_{-}}+c\rho+e\, ,\hspace{0.5cm}p_\pm=  \dfrac{lm^{2}+2\mu\pm\sqrt{m^{4}l^{2}+4m^{2}\mu^{2}l^{2}}}{4\mu} \, .
\end{equation}
where we have chosen $a_\pm, c$ and $e$ as integration constants.  Although these can be arbitrary functions of the null-direction $u$ \cite{vacuaEMG}, leading to radiative spacetimes,  to compute the energy content of the spacetime as an integral at the spatial infinity, we shall concentrate on the ``static" AdS wave case. Note that these spacetimes do not have a constant Riemann curvature, and we will use as a constant curvature background the metric obtained by setting $ F(\rho)=0 $ in the metric (\ref{adswave}). 

Considering $ \bar{\xi}^{\mu}=\delta^{\mu}{}_{t} $, the total energy reads
\begin{align}\label{massadswave}
E  &= Q^{\mu}(\bar{\xi}) =2e-\frac{2e}{l\mu}-\frac{2e}{l^{2}m^{2}}+2\left(  \frac{(2p_{+}-1)^{2}%
}{l^{2}m^{2}}-1-\frac{(2p_{+}-1)}{l\mu}\right)  a_{+}(p_{+}-1)\rho_{c}^{p_{+}%
}\nonumber\\
& +2\left(\frac{(2p_{-}-1)^{2}}{l^{2}m^{2}}-1-\frac{(2p_{-}-1)}{l\mu
}\right)a_{-}(p_{-}-1)\rho_{c}^{p_{+}}\, ,%
\end{align}
where we have introduced a cut-off $\rho_c$ that has to be taken to infinity. Finiteness of the mass therefore requires the couplings to be such that $p_+$ and $p_-$ are negative  (one of them could also vanish, but this leads to logarithmic terms having slower fall-off to AdS than that considered in this section). The energy of this ``static"-AdS wave solution  is then given by the first three terms in (\ref{massadswave}). 

With respect to the axial symmetry of the background, we use the Killing vector $ \bar{\xi}^{\mu}=\delta^{\mu}{}_{\phi} $, interestingly leading to the conserved charge
\begin{equation}
J = Q^{\mu}(\bar{\xi})  =El 
\end{equation}
with $E$ given in equation (\ref{massadswave}).
\newpage 

\section{Mass and angular momentum for axisymmetric vacua of log-EMG}

\textbf{Log deformations of Extremal BTZ geometry}
\\

  When the chirality relation $\mu={\dfrac {l{m}^{2}}{1-{m}^{2}{l}^{2}}} $ is fulfilled, EMG admits asymptotically AdS solutions with logarithmic deformations \cite{vacuaEMG}. The same occurs in Chiral Gravity \cite{Maloney:2009ck} as well as in Minimal Log Gravity \cite{Giribet:2014wla}. Remarkably, as in those cases, even though the asymptotic behaviour is relaxed with respect to the Brown-Henneaux \cite{BrownHenneaux}, the charges are finite. The potentially divergent terms are multiplied by the chiral relation, which identically vanishes on-shell. 
The background in this case, is the extremal rotating BTZ geometry, which is also locally AdS, but globally differs from the maximally symmetric AdS$_3$ spacetime \cite{BTZHenneaux}.

At the chiral point, the following metric solves the field equations of EMG
\begin{equation}
ds^{2}=-Ndt^{2}+\dfrac{dr^{2}}{N}+r^{2}(N_{\phi}dt+d\phi)^{2}+N_{k}(dt+ld\phi)^{2}
\end{equation} 
where
\begin{equation}
N=-M+\dfrac{r^{2}}{l^{2}}+\dfrac{M^{2}l^{2}}{4r^{2}}\hspace{0.5cm},\hspace{0.5cm}N_{\phi}=\dfrac{Ml}{2r^{2}}\hspace{0.5cm},\hspace{0.5cm}N_{k}=k \ln\left(r^{2}-\dfrac{Ml^{2}}{2}\right)  
\end{equation}
and 
$M = k = 0$ defines the background metric (the massless BTZ geometry).  

For generic couplings  we find from \eqref{chargesfinal} that the mass and the angular momentum are respectively given by 
\begin{align*}
E  & =\frac{2k(5\mu+4l\mu\mu_{c}-3\mu_{c})}{\left(  1+l\mu_{c}\right)
l^{2}\mu m^{2}}+\frac{\left(  \mu_{c}-\mu\right)  \left(  4k\ln\left(
r\right)  +M\right)  }{l^{2}m^{2}\mu\left(  1+l\mu_{c}\right)  }\ ,\\
J  & =-\frac{2k(5\mu+4l\mu\mu_{c}-3\mu_{c})}{\left(  1+l\mu_{c}\right)  l\mu
m^{2}}+\frac{(\mu-\mu_{c})\left(  4k\ln\left(  r\right)  +M\right)  }{\left(
1+l\mu_{c}\right)  l\mu m^{2}} 
\end{align*}
where the chiral value of $\mu$ is $\mu_c=m^2l/(1-m^2l^2)$.  It is clear  that
\begin{equation}
E = 4k+\frac{4k}{l^{2}m^{2}}  \qquad  J = - l E
\end{equation}
 at the chiral point $\mu=\mu_c$.

\section{Conclusion}\label{sec4}

In this note we have constructed the general formula for computing conserved charges of solutions to the Exotic Massive Gravity theory, recently introduced in \cite{EMG}. The theory is the next-to-simplest gravitational theory fulfilling the so-called ``third-way consistency" \cite{Bergshoeff:2015zga}. The theory does not admit a Lagrangian formulation purely given in terms of the metric, nevertheless, since the divergence of the field equations is proportional to the same field equations, the theory turns out to be consistent. Using the Abbot-Deser-Tekin approach to compute conserved currents for the theory associated with Killing vectors of the background, we showed  that this leads to finite charges, even at the chiral point at which the solutions acquired a relaxed asymptotic behaviour containing log-terms.

As described in \cite{EMG},   theories fulfilling  ``third-way" consistency are constructed from a seed symmetric, rank-two tensor with an identically vanishing divergence. A simple candidate for these seed tensors are the Euler-Lagrange derivatives with respect to the metric, of diffeomorphism invariant actions, even though it remains an open problem whether this procedure leads to the most general tensor fulfilling such properties, see \cite{Deser:2018mje}. Following this approach a general ``third-way consistent" theory containing a seed that is the field equations of a generic quadratic action was recently constructed \cite{bachian}. It would be interesting to explore further properties of the whole, infinite hierarchy of such theories in the future.

\bigskip

{\bf Acknowledgements}\\
S. N. Sajadi acknowledge the support of Shahid Beheshti University.  This work was supported in part by the Natural Sciences and Engineering Research Council of Canada. J.O. appreciates the support of CONICYT through the grant FONDECYT Regular 1181047.

\end{document}